\documentclass[a4paper,12pt]{article}
\usepackage[a4paper,text={16.8cm,22.4cm}]{geometry}
\usepackage{amsmath,amssymb,bm,graphicx,color,psfrag,booktabs,textcomp,multirow}
\usepackage[normalem]{ulem}
\usepackage{cancel}
\allowdisplaybreaks 
\def\rd{\mathrm{d}}

\begin{document}

\begin{titlepage}

\begin{flushright}
July 15, 2014 \\
OUTP-14-13P \\
\end{flushright}

\vspace{0.7cm}
\begin{center}
\Large\bf
\boldmath
The transverse-momentum spectrum of Higgs bosons \\ near threshold at NNLO
\unboldmath
\end{center}

\vspace{0.2cm}
\begin{center}
{\sc Thomas Becher$^a$, Guido Bell$^b$, Christian Lorentzen$^a$ and Stefanie Marti$^a$}\\
\vspace{0.4cm}
{\sl 
 ${}^a$\,Albert Einstein Center for Fundamental Physics\\
Institute for Theoretical Physics \\
University of Bern\\
Sidlerstrasse 5, 3012 Bern, Switzerland\\[0.4cm]

${}^b$\,Rudolf Peierls Centre for Theoretical Physics\\ 
University of Oxford\\ 
1 Keble Road, Oxford OX1 3NP, United Kingdom}
\end{center}

\vspace{1.0cm}
\begin{abstract}
We give next-to-next-to-leading order (NNLO) predictions for the Higgs production cross section at large transverse momentum in the threshold limit. Near the partonic threshold, all radiation is either soft or collinear to the final state jet which recoils against the Higgs boson. We find that the real emission corrections are of moderate size, but that the virtual corrections are large. We discuss the origin of these corrections and give numerical predictions for the transverse-momentum spectrum. The threshold result is matched to the known NLO result and implemented in the public code {\sc PeTeR}.
\vspace{0.2cm}
\noindent 
\end{abstract}

\vfil

\end{titlepage}

\section{Introduction}

Precision studies of Higgs properties are a central part of the physics program at the CERN Large Hadron Collider (LHC). The second LHC run at higher center-of-mass energy will allow to measure kinematic distributions of Higgs bosons such as the transverse-momentum spectrum. Knowledge of the spectrum is important when dealing with backgrounds to Higgs production, but it can also be used to search for the effects of new physics. In the past, studies of this type were mostly concerned with light particles at low to intermediate $p_T$ values, but more recently several papers have investigated the possibility to use information on the shape of the spectrum at $p_T$ values larger than the top-quark mass \cite{Banfi:2013yoa,Azatov:2013xha,Grojean:2013nya,Schlaffer:2014osa}. For such values, finite top-quark mass effects become relevant, and it might be possible to disentangle the top-quark contribution from the effects of new heavy particles coupling to the Higgs boson.

On the theory side, Higgs physics is challenging, because Higgs cross sections suffer from large perturbative corrections, so that higher-order contributions are needed to achieve reliable theoretical predictions. For the total cross section, there are ongoing efforts to compute the fourth-order terms in the perturbative expansion. As an important first step towards the full next-to-next-to-next-to-leading order (N$^3$LO) result, the N$^3$LO terms have recently been computed in the threshold limit \cite{Anastasiou:2014vaa}. At non-zero transverse momentum $p_T$ of the Higgs boson, on the other hand, the cross section is currently only known to NLO \cite{deFlorian:1999zd,Ravindran:2002dc,Glosser:2002gm}, with ongoing efforts to extend the result to NNLO. For the dominant, purely gluonic partonic channel, first NNLO results were obtained about a year ago in \cite{Boughezal:2013uia}, and updated, preliminary results were presented at a recent conference \cite{loopfest13}. In the present paper, we compute the rate for Higgs production at non-zero transverse momentum $p_T$ to NNLO in the threshold limit. At the partonic level, the threshold cross section consists of all the singular distributions. These yield the dominant part of the hadronic cross section, in particular at large transverse momentum, where the contribution from regular terms is suppressed by the fall-off of the parton distribution functions (PDFs).

We recently presented all ingredients to perform threshold resummation at next-to-next-to-next-to-leading logarithmic (N$^3$LL) accuracy for electroweak boson production at large transverse momentum \cite{Becher:2013vva}. At this accuracy, the resummed result includes the full NNLO threshold cross section. Near threshold, the electroweak boson recoils against a low-mass jet and the partonic cross section factorizes into a hard function, a jet function and a soft function. For the channel $a+b \to H+ j_c $, the factorization formula takes the form
\begin{equation} \label{eq:factorization}
\hat{s} \frac{\rd\hat{\sigma} }{\rd\hat{u}\, \rd\hat{t}} = H_{ab} (\hat{u},\hat{t}) \, (J_c \otimes S_{ab})(m_X^2)\, ,
\end{equation} 
where the partonic Mandelstam variables are $\hat{s} = (p_a+p_b)^2$, $\hat{t} = (p_a-q)^2$ and $\hat{u} = (p_b-q)^2$, with $q$ the Higgs boson momentum, and $q^2=M_H^2$. The hard function $H_{ab} $ captures the purely virtual corrections to the hard scattering process, while the jet and soft functions $J_c$ and $S_{ab}$ describe the real emissions, which can either be collinear to the final state jet or soft. The convolution of the jet and soft functions depends on the invariant mass of the partonic final state jet $m_X$, which goes to zero in the threshold limit. The jet and soft functions were computed to two-loop order earlier in \cite{Becher:2006qw,Becher:2010pd} and \cite{Becher:2012za}. In our recent paper \cite{Becher:2013vva}, we extracted the final ingredient for N$^3$LL resummation, namely the two-loop hard function, from the results for the two-loop helicity amplitudes for theses processes \cite{Gehrmann:2013vga,Gehrmann:2011aa}. 

Our results for $W$ and $Z$ production have been implemented into a public code {\sc PeTeR} \cite{peter}. In the meantime, we have also implemented the resummation as well as the NLO result for Higgs production into a new release of this code, and we are now in the position to present numerical results also in this case. For vector bosons, the two-loop corrections turned out to be moderate, but in contrast we find very large corrections for Higgs production. These corrections are due to large higher-order terms in the hard function, and they significantly change the results from threshold resummation at lower precision. For the Higgs transverse-momentum spectrum, threshold resummation was first performed at NLL accuracy in \cite{de Florian:2005rr} and it was found that NLL effects increase the NLO cross section by about 10\% and reduce the scale dependence by a factor of two. At NLL accuracy, only the tree-level hard function is included. Very recently, the resummation was performed to NNLL accuracy, which includes the one-loop hard function \cite{Huang:2014mca}. The authors find that NNLL resummation reduces the NLO result by about 10\% and the scale dependence by more than a factor of two. In contrast, after computing the full NNLO threshold result, we find a significant increase in the cross section, as large as 50\% over the NLO result. The source of this increase are large positive two-loop corrections to the hard function which only enter at N$^3$LL accuracy.

In Section \ref{size}, we analyze the two-loop corrections to the threshold cross section in detail and suggest a way to improve the perturbative convergence of the hard function using renormalization group methods. We also determine the appropriate scale choices for the different ingredients in the factorization formula. Based on these results, we give numerical predictions for the cross section at large transverse momentum in Section \ref{numer}. Our NNLO results are valid in the large-$m_t$ limit, but we also discuss finite top mass effects which are known at LO in Section \ref{numer}.

\section{Size of the perturbative corrections\label{size}}

One advantage of the effective theory framework \cite{Bauer:2000yr,Bauer:2001yt,Beneke:2002ph} we use here is that we can evaluate each part of the factorization formula at its natural renormalization scale, which should be chosen to avoid large logarithmic corrections. Using renormalization group (RG) techniques, the ingredients are then evolved to a common scale $\mu_f$ at which the PDFs are evaluated. For the hard function $H_{ab} (\hat{u},\hat{t},\mu)$, one expects the natural value of the scale $\mu$ to be of the order of the transverse momentum $p_T$. In order to combine the hard function with the remaining cross section, one can solve the RG evolution equation for this function, which yields
\begin{equation} \label{rgsol}
H_{ab} (\hat{u},\hat{t},\mu) = U(\mu_h, \mu)\, H_{ab} (\hat{u},\hat{t},\mu_h)\,.
\end{equation}
The evolution factor $U(\mu_h, \mu)$ depends on the anomalous dimensions of the hard function. The construction of the hard function from the results for the four-point helicity amplitudes \cite{Gehrmann:2013vga,Gehrmann:2011aa} is discussed in detail in \cite{Becher:2013vva}. It is obtained by squaring renormalized helicity amplitudes,
\begin{equation}
H_{ab} (\hat{u},\hat{t},\mu)  = \sum |\mathcal{M}_{ab}(\hat{u},\hat{t},\mu) |^2\,.
\end{equation}
The sum indicates that one sums (averages) over outgoing (incoming) colors and helicities of the particles. Because it will be relevant for our discussion below, we give the RG evolution equation for the $gg\to H g$ amplitude. Due to factorization constraints \cite{Becher:2009cu,Gardi:2009qi,Becher:2009qa,Dixon:2009ur}, it has the form
\begin{equation}\label{gammaH}
\frac{{\rm d}}{{\rm d}\ln \mu} \mathcal{M}_{gg} (\hat{u},\hat{t},\mu) = \left[ \frac{C_A}{2} \gamma_{\rm cusp}(\alpha_s)\,  \left(\ln \frac{-\hat{s}}{\mu^2}+ \ln\frac{-\hat{t}}{\mu^2}+\ln\frac{-\hat{u}}{\mu^2} \right) + 3 \gamma_g(\alpha_s) \right] \, \mathcal{M}_{gg} (\hat{u},\hat{t},\mu)\,,
\end{equation}
at least up to three-loop accuracy. Explicit three-loop results for the anomalous dimensions $\gamma_{\rm cusp}$ and $\gamma_{g}$ can be found in the appendix of \cite{Becher:2009qa}. 

\begin{figure}[t!]
\centering
\begin{tabular}{c}
\includegraphics[width=0.87\textwidth]{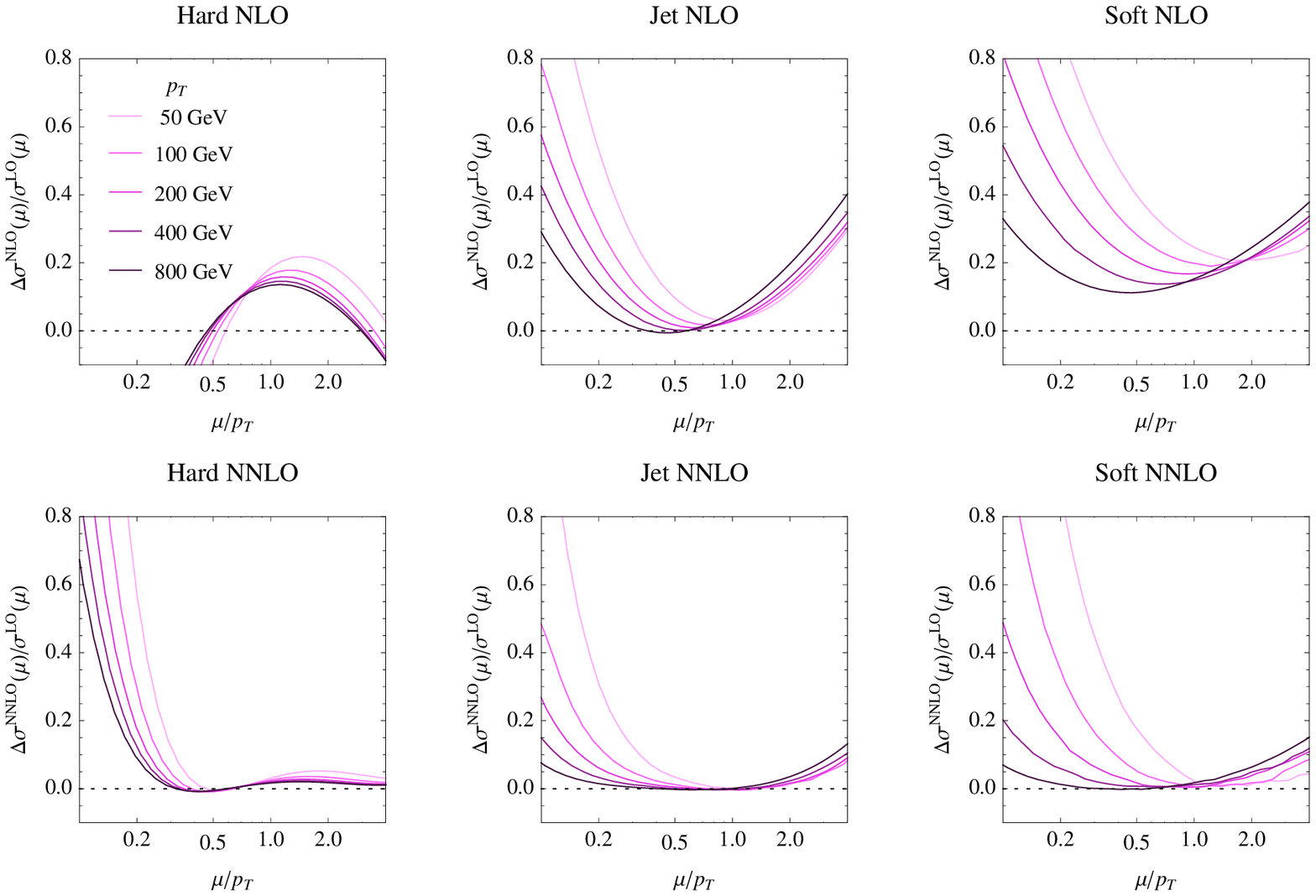}
\end{tabular}
\vspace{-2.5mm}
\caption{Size of the corrections to the hard, jet, and soft function for $Z$-production.
\label{fig:hjsZ}}
\vspace{7mm}
\begin{tabular}{c}
 \includegraphics[width=0.87\textwidth]{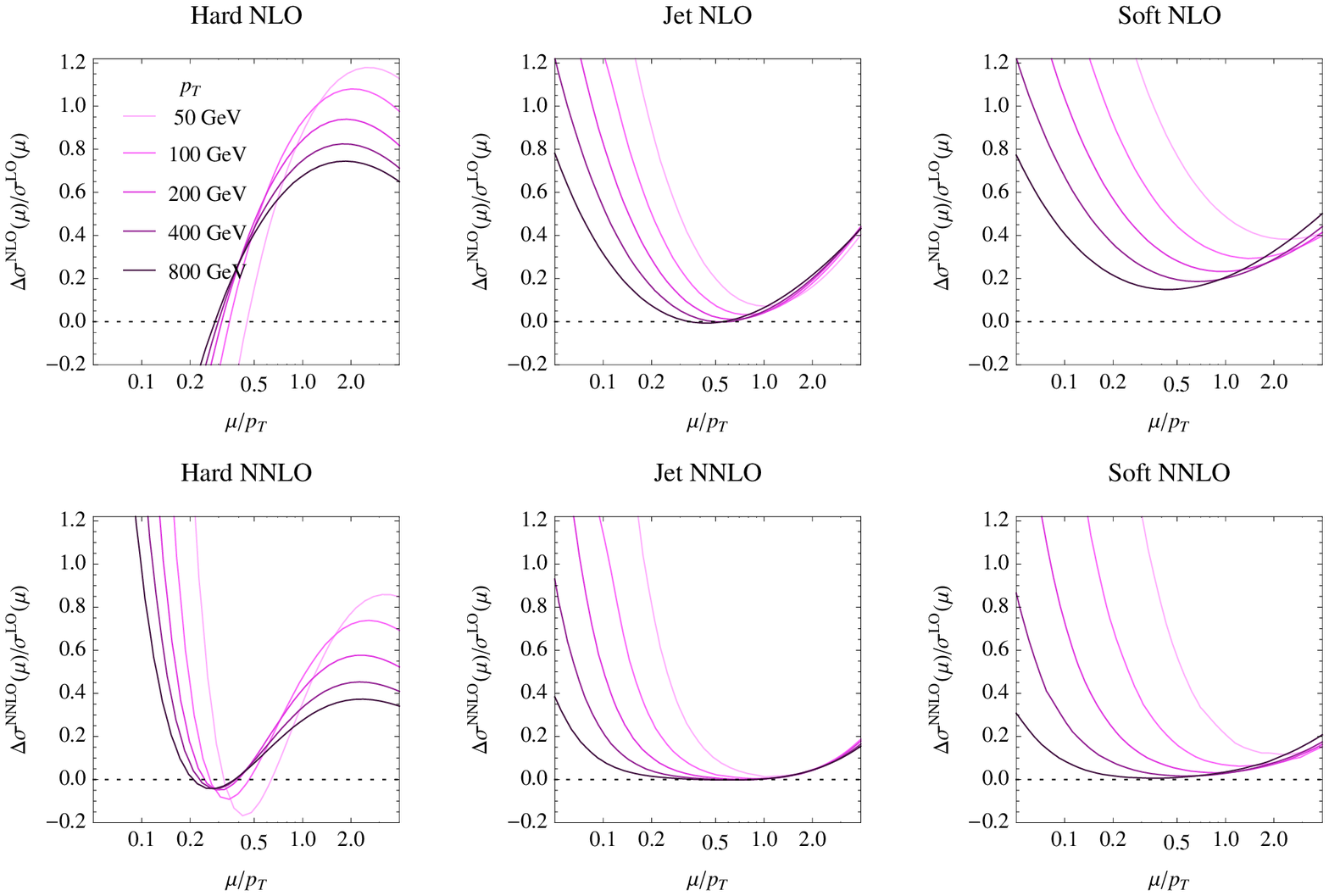} 
\end{tabular}
\vspace{-2.5mm}
\caption{Size of the corrections to the hard, jet, and soft function for Higgs production.
\label{fig:hjsH}}
\end{figure}

The solution \eqref{rgsol} provides a representation of the hard function which is free of large perturbative logarithms as long as the starting scale $\mu_h$ of the RG evolution is chosen properly. Similarly, one can obtain RG-improved versions of the jet and soft functions. For these functions, it is not immediately clear what one should choose as an appropriate scale. While $\mu_j = m_X$ is a natural choice at the partonic level, the invariant mass $m_X$ is integrated over a range from $m_X=0$ at the threshold up to large values when the convolution with the PDFs is evaluated. For the hadronic cross section, we would like to use an average value $\langle m_X \rangle$ as our choice of $\mu_j$. The  value of $\langle m_X \rangle$ will depend on the shape of the PDFs and can in general only be determined numerically. Detailed studies of the size of the hard, jet and soft corrections for $W$ and $Z$ production were performed in \cite{Becher:2011fc,Becher:2012xr} based on the method of \cite{Becher:2007ty}. An interesting alternative method to choose the proper scales was proposed recently in \cite{Sterman:2013nya}. It determines the scale from PDF luminosities and cannot immediately be applied in our case. However, for  inclusive Higgs production, it leads to  similar numerical results to the method we adopt here. 

The Higgs transverse-momentum spectrum has some interesting similarities to the $Z$-boson results, but also shows a dramatic difference that we now examine in detail. To this end, we show in Figures \ref{fig:hjsZ} and \ref{fig:hjsH} the size of the individual corrections to the $Z$ and $H$ cross sections. In each of the plots, we only switch on one individual correction, either to the hard, the jet or the soft function, and we study its size as a function of the renormalization scale. Since we are interested in individual corrections, we do not perform any resummation at this stage and use a common value for all the scales, i.e.\ we set $\mu=\mu_h=\mu_j=\mu_s=\mu_f$.  Dividing by the LO cross section, the individual one and two-loop corrections have the form
\begin{align}\label{formcorr}
\Delta \sigma^{\rm NLO}(\mu) / \sigma^{\rm LO}(\mu) &= \alpha_s(\mu) (c_2 L^2 + c_1 L + c_0)\,, \\
\Delta \sigma^{\rm NNLO}(\mu) / \sigma^{\rm LO}(\mu) &= \alpha_s^2(\mu) (d_4 L^4 + d_3 L^3 + d_2 L^2 +d_1 L +d_0 )\,, \nonumber
\end{align}
where $L = \ln \mu/ \Lambda$. The scale $\Lambda \sim p_T$ for the hard function, and $\Lambda \sim \langle m_X \rangle$ for the jet function.  For the soft function $\Lambda \sim\langle E_s \rangle$, the average energy of the soft radiation. Looking at the scale dependence of the corrections  allows us to choose a proper value of the scale: if we choose the scale too low or too high, we end up with large corrections due to the Sudakov logarithms in \eqref{formcorr}. The logarithmic plots in Figures \ref{fig:hjsZ} and \ref{fig:hjsH} nicely display the second-order (fourth-order) polynomial form of the NLO (NNLO) corrections.

Looking at the corrections to $Z$-production, we find that the proper scale choice for the hard function is indeed $\mu_h\sim p_T$. The scale of the jet and soft functions is lower, but not dramatically lower than $p_T$. This implies that there are no large scale hierarchies in the cross section. The resummation of logarithms should therefore only be a moderate effect. This observation was made earlier in \cite{Becher:2013vva,Becher:2011fc,Becher:2012xr}, where it was found that resummation has a small effect on the central value but leads to somewhat reduced scale uncertainties. One sees from the plots that the scales determined from the NLO and NNLO corrections are almost identical, as they should be if there is a natural scale associated with the corrections. What is also obvious from the plots is that all of the NNLO corrections to $Z$-production are small as long as the scales are chosen properly. Indeed, we found in \cite{Becher:2013vva} that the two-loop corrections to $W$ and $Z$ production are moderate, of the order of 5\%.

\begin{figure}[t!]
\centering
\begin{tabular}{cc}
\raisebox{0.40cm}{\includegraphics[width=0.393\textwidth]{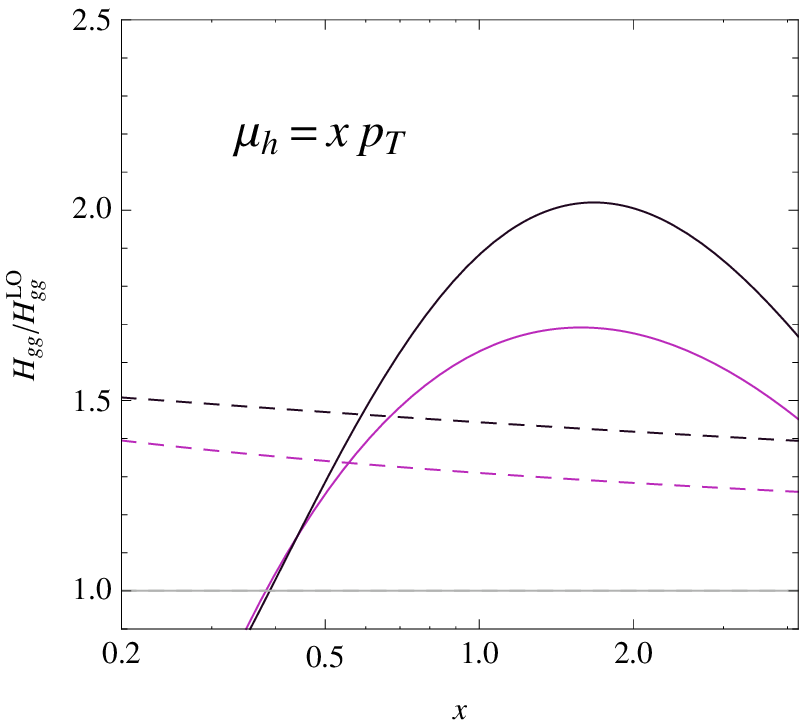}} & 
\includegraphics[width=0.4\textwidth]{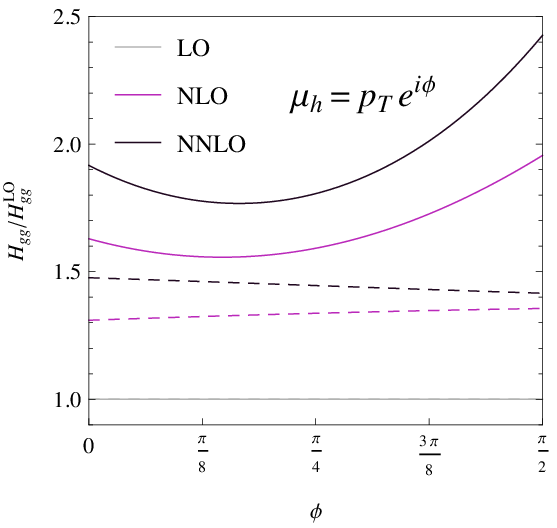} 
\end{tabular}
\vspace{-0.4cm}
\caption{Size of the corrections to the hard function for real and complex $\mu_h$. The results are for $p_T=0.2\,{\rm TeV}$ and $\hat{s}=(0.5\,{\rm TeV})^2$. The solid lines show the hard function $H_{ gg} (\hat{u},\hat{t},\mu_h)$, while the dashed lines show the result for the reduced hard function $\widetilde{H}_{ gg}(\hat{u},\hat{t})$.\label{complmu}}
\end{figure}

Let us now contrast this situation with the one in Higgs production shown in Figure \ref{fig:hjsH}. First of all, one observes that the corrections to the jet and soft functions as well as the associated scales are quite similar to the $Z$-boson case. This is not an accident, but simply a reflection of the fact that the same jet and soft functions are relevant for both processes. This is clear for the gluon and quark jet functions, which obviously arise in both cases, but it is also true for the soft function. In fact, the two-loop soft functions in the different partonic channels only differ by their color factor, which is $C_F-C_A/2$ for $q\bar q \to g$ and $C_A/2$ for $q g \to q$ and $g g \to g$ \cite{Becher:2012za}. However, whereas the same jet and soft functions are involved in both cases, the hard function for Higgs production is dramatically different. One observes very large corrections, of order  100\% at NLO and 50\% at NNLO, even for natural scale choices $\mu_h \sim p_T$. 

This pattern of large corrections is familiar from the total Higgs production cross section. Also in this case one encounters very large virtual corrections, even for the seemingly natural choice of the hard scale $\mu_h^2 =\hat{s}$. For the total cross section, the hard function is given by the square of the scalar form factor and the large corrections could be traced back to the analytic continuation of the space-like form factor to time-like kinematics \cite{Ahrens:2008qu,Ahrens:2008nc}. The analytic continuation of Sudakov double logarithms $\alpha_s \ln^{2}(-\hat{s}/\mu_h^2)$ produces $\pi^2$ terms which due to the associated color factor give large corrections to the cross section. Since these terms are tied to Sudakov logarithms, they can be resummed as was observed a long time ago \cite{Parisi:1979xd,Magnea:1990zb}. A simple way of achieving this resummation for the total cross section is to choose a time-like value of the hard scale $\mu_h^2 = - \hat{s}$. For this choice the Sudakov logarithms in the expansion are minimized and the $\pi^2$ terms are resummed by RG evolution from $\mu_h^2= - \hat{s}$ back to positive values of $\mu_h^2$ \cite{Ahrens:2008qu,Ahrens:2008nc}. Unfortunately, the same procedure cannot immediately be applied to the hard function with a jet in the final state, relevant for the Higgs transverse-momentum spectrum. As is obvious from equation \eqref{gammaH}, the hard function contains in this case double logarithms in $\hat{s}$, $\hat{t}$ and $\hat{u}$ and there will be imaginary parts for any value of $\mu_h^2$. Indeed, plotting the hard function as a function of $\mu_h = p_T\, e^{i\varphi}$, one finds that the corrections are roughly of the same size, no matter what value of $\varphi$ is chosen, as was observed in \cite{Jouttenus:2013hs} and can be seen in Figure \ref{complmu}. Note that $\alpha_s(\mu_h)$ and the amplitudes are functions of $\mu_h^2$; it is thus sufficient to consider $|\varphi| < \pi/2$. In the plot we show the result for positive arguments $\varphi$. The values at negative $\varphi$ are very similar.

A simple procedure to address the problem of large corrections exploits the fact that the anomalous dimension of the $Hgg$ scalar form factor $F_S(\hat{s},\mu)$ and the $gg\to H g$ amplitude are closely related. The RG equation for the form factor reads
\begin{equation}\label{gammaF}
\frac{{\rm d}}{{\rm d}\ln \mu} F_S(\hat{s},\mu) = \left[ C_A\, \gamma_{\rm cusp}(\alpha_s)\,\ln \frac{-\hat{s}}{\mu^2} + 2\gamma_g(\alpha_s)\right]  \, F_S(\hat{s},\mu) \,.
 \end{equation}
The form factor is $F_S=\alpha_s \,C_t\, C_S$ and $C_t$ and $C_S$ are given explicitly in \cite{Ahrens:2008nc}.
If one defines a reduced amplitude as
\begin{equation}\label{Mred}
\widetilde{\mathcal M}_{gg} (\hat{u},\hat{t}) = \frac{{\mathcal M}_{gg} (\hat{u},\hat{t},\mu)}{\sqrt{F_S(\hat{s},\mu)  F_S(\hat{t},\mu)  F_S(\hat{u},\mu) }}\,,
 \end{equation}
this amplitude will be independent of the scale $\mu$ and one can then use the RG equation \eqref{gammaF} to resum large corrections to the individual form factors in \eqref{Mred}. However, such an approach may be overly simplistic. The problem is that the reduced function is still a function of two variables, so it can contain terms of the form $\alpha_s \ln^2 \hat{t}/\hat{s}$ which can give rise to large corrections. In particular, at small transverse momentum the amplitude $\mathcal M_{gg}(\hat{u},\hat{t},\mu)$ factorizes into a form factor $F_S(\hat{s},\mu)$ times a $g\to gg$ splitting amplitude. It is clear that the reduced amplitude \eqref{Mred} will not capture all large corrections in this region.

\begin{figure}[t!]
\centering
\includegraphics[width=0.5\textwidth]{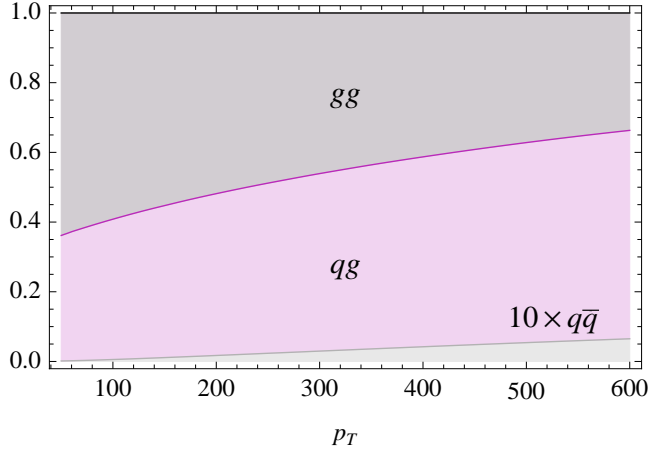} 
\vspace{-0.2cm}
\caption{Relative contribution of different partonic channels to the NNLO correction for the default scale choice $\mu= p_T$. The $qg$ contribution includes all partonic channels with a single \mbox{(anti-)quark} in the initial state.
\label{fig:channels}}
\end{figure}    

Let us discuss the numerical effects of the above prescription. To this end, we choose a generic phase-space point with $\hat{s}=1\, {\rm TeV}^2$, $\hat{t}=-0.4\, {\rm TeV}^2$ and $M_H=0.1\, {\rm TeV}$. These values imply that the transverse momentum is $p_T^2 = \hat{t}\hat{u}/\hat{s} \approx ( 0.5\, {\rm TeV})^2$. For the renormalization scale, we use $\mu = 0.6\, {\rm TeV}$ and obtain 
  \begin{align}
H_{gg} (\hat{u},\hat{t},\mu) &=H_{gg}^{\rm LO} (\hat{u},\hat{t},\mu)\left( 1+7.77234\, \alpha _s+38.2661 \,\alpha _s^2 \right)\,, \\
\widetilde{H}_{ gg} (\hat{u},\hat{t}) &=\widetilde{H}_{ gg}^{\rm LO} (\hat{u},\hat{t}) \left( 1+1.92209\, \alpha _s+8.29574 \,  \alpha _s^2 \right)\,. \nonumber
 \end{align}
We work at the same kinematic point considered in our previous paper \cite{Becher:2013vva}, but the above numbers include the corrections to the Wilson coefficient $C_t$ of the effective $Hgg$ operator obtained after integrating out the top quark. We find that the corrections are significantly reduced both at NLO and NNLO. For a different phase-space point, the reduction can also be seen by comparing the dashed to the solid lines in Figure \ref{complmu}. 
 
Since $\hat{t}$ and $\hat{u}$ are negative, the associated form factors in \eqref{Mred} do not suffer from large perturbative corrections and only the form factor $F_S(\hat{s},\mu)$ needs to be RG improved. One can thus  simply multiply the cross section by a prefactor to improve the convergence,
 \begin{equation}\label{eq:impr}
\left(\frac{d\sigma}{dp_T}\right)^{\rm impr.} =  \left|\frac{F_S(p_T^2,\mu_h) U_S(\mu_h, \mu)}{F_S(p_T^2,\mu)}\right|\, \frac{d\sigma}{dp_T} \,.
 \end{equation}
When improving the hadronic cross section, we can evaluate the form factor at the typical momentum transfer $Q^2=p_T^2$ instead of the scale $Q^2=\hat{s}$ which arises at the partonic level. Choosing $\mu_h = i p_T$ gives a well-behaved perturbative expansion in the numerator, and the denominator divides out the large corrections to the cross section. The RG-evolution factor $U_S(\mu_h, \mu)$, whose explicit form can be found in \cite{Ahrens:2008qu,Ahrens:2008nc}, then resums the large corrections. 

We can apply the same improvement also to the other partonic channels, which involve quarks. In this case, we need to multiply the amplitudes with an appropriate combination of vector and scalar form factors. For the $q g \to H q$ channel, for example, the relevant combination is
 \begin{equation}\label{Mred2}
 \widetilde{\mathcal M}_{qg} (\hat{u},\hat{t}) = \frac{\sqrt{F_S(\hat{u},\mu)} }{ \sqrt{F_S(\hat{s},\mu) F_S(\hat{t},\mu) } F_V(\hat{u},\mu) }\, {\mathcal M}_{qg} (\hat{u},\hat{t},\mu)\,.
 \end{equation}
The reason for the difference to \eqref{Mred} is that the $u$-channel logarithm in \eqref{gammaH} now has a color factor of $C_F - C_A/2$, whereas the color factor associated with the vector form factor is $C_F$. For this channel, the corrections are
\begin{align}
H_{qg} (\hat{u},\hat{t},\mu) &=H_{qg}^{\rm LO} (\hat{u},\hat{t},\mu)\left( 1+8.38935\, \alpha _s +40.0591 \, \alpha _s^2 \right)\,, \\
\widetilde{H}_{q g} (\hat{u},\hat{t}) &=\widetilde{H}_{q g}^{\rm LO} (\hat{u},\hat{t}) \left( 1 + 6.04455  \,\alpha_s + 23.4922  \,\alpha _s^2 \right)\,, \nonumber
\end{align}
and in the $q \bar{q}$ channel one obtains
\begin{align}
H_{q\bar{q}} (\hat{u},\hat{t},\mu) &=H_{q\bar{q}}^{\rm LO} (\hat{u},\hat{t},\mu)\left( 1+3.60093\, \alpha _s+14.8465\, \alpha _s^2 \right)\,, \\
\widetilde{H}_{q\bar{q}} (\hat{u},\hat{t}) &=\widetilde{H}_{q\bar{q}}^{\rm LO} (\hat{u},\hat{t}) \left( 1+3.32609 \,\alpha _s+11.6103  \,\alpha _s^2 \right)\,. \nonumber
\end{align}
The size of the corrections is reduced, but not as much as in the $gg$ channel. The relative NNLO contribution of the individual channels to the cross section is shown in Figure \ref{fig:channels}. For low $p_T$, the $gg$ channel yields the dominant contribution to the cross section. The contribution of the $q\bar{q}$ channel is numerically negligible, but the $qg$ channels contribute a significant fraction of the cross section. In fact, for $p_T  \gtrsim 250\,{\rm GeV}$ they give the dominant contribution. Since the RG improvement only affects the $s$-channel form factor and the dependence of the reduced amplitudes \eqref{Mred} and \eqref{Mred2} on this form factor is the same, it follows that the prescription \eqref{eq:impr} is relevant for both the $gg$ and $qg$ channels. Given that the $q\bar{q}$ channel is negligible, it is therefore appropriate to use \eqref{eq:impr} for the full cross section.

\begin{figure}[t!]
\centering
\begin{tabular}{ccc}
\includegraphics[width=0.31\textwidth]{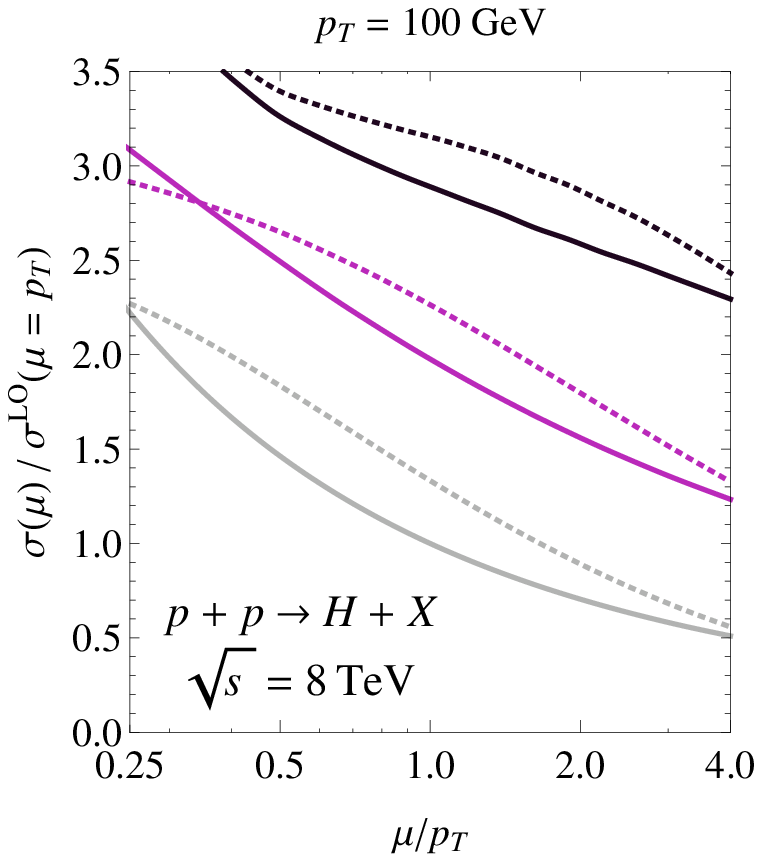} & \includegraphics[width=0.31\textwidth]{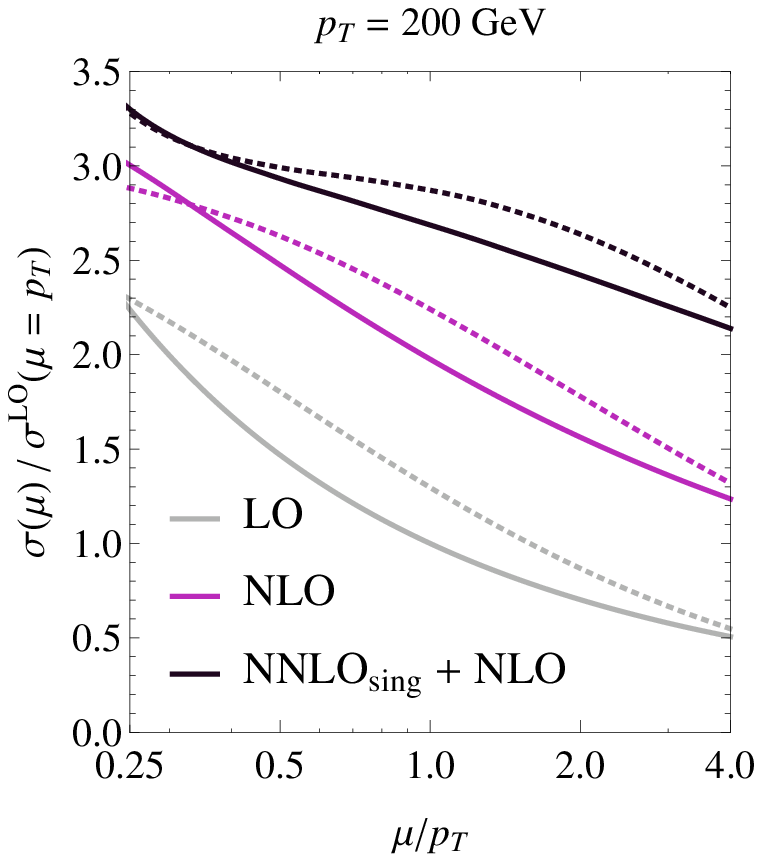} & \includegraphics[width=0.31\textwidth]{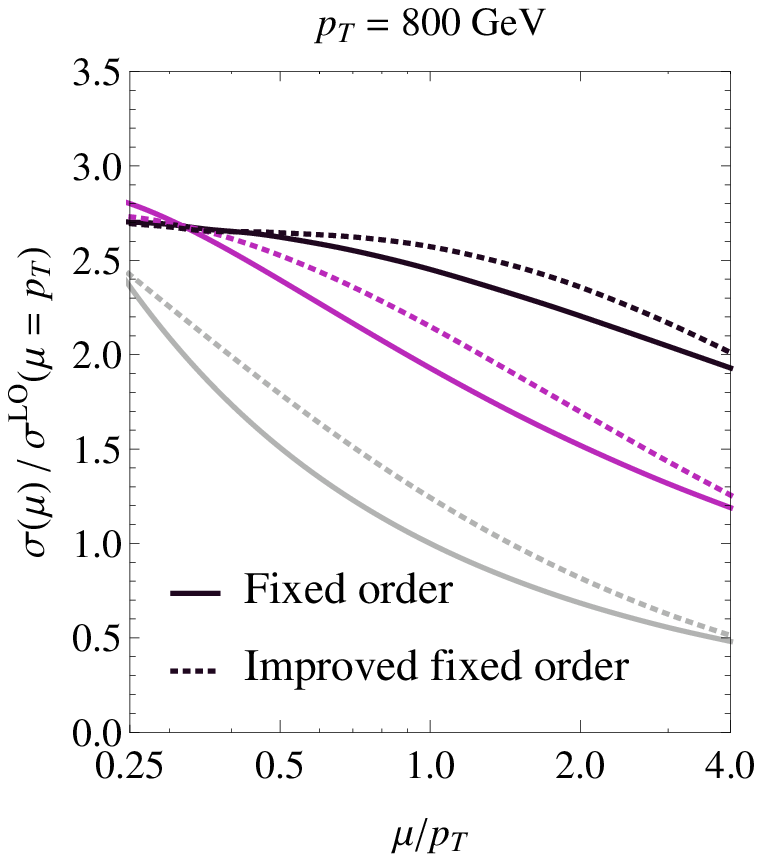} 
\end{tabular}
\caption{Scale dependence of the cross section at LO (gray), NLO (purple) and NNLO$_{\rm sing} +$NLO (black). The dashed lines show the result with RG improvement according to the prescription \eqref{eq:impr}. 
\label{fig:scale}}
\end{figure}

In Figure \ref{fig:scale}, we show the scale dependence of the cross section at different orders in the perturbative expansion. In these plots, we set the hard, jet and soft scales to a common value, $\mu = \mu_h=\mu_j= \mu_s$, and we also set the factorization scale $\mu_f=\mu$. If all scales are set equal, the resummation is switched off and we obtain the fixed-order result for the threshold terms. To distinguish these from the full result, we denote them by N$^n$LO$_{\rm sing}$ since they consist of singular distributions whose explicit form is given in \cite{Becher:2013vva}. At N$^3$LL, we obtain threshold terms up to NNLO$_{\rm sing}$. For our most accurate result, denoted by NNLO$_{\rm sing}+$NLO, the threshold terms are then matched to the full NLO result. The figure clearly shows that the higher-order corrections are large, and that the convergence is 
only slightly improved for very high values of $p_T$. In all our plots, we use NNLO PDFs. The corrections would look smaller if we had used LO PDFs for the lowest-order cross section because of the associated larger value of $\alpha_s$. However, our goal here is to assess the size of the perturbative corrections, and to this end it is more informative to keep the PDFs and $\alpha_s$ fixed. In Figure \ref{fig:scale}, we also give the result for the improved cross section according to our ansatz \eqref{eq:impr}, shown by the dashed lines. We find that the improvement is only moderate at the level of the cross section, despite the fact that both the reduced hard function $\widetilde{H}_{gg}$ and the improved scalar form factor $F_S$ have well-behaved perturbative expansions. As can be read off from Figure \ref{complmu}, the corrections to the reduced amplitude are about 35\% at NLO and 15\% at NNLO. For $p_T= 200\, {\rm GeV}$ and default scale choices $\mu_h = i p_T$ and $\mu = p_T$, the expansion of the form factor takes the form 
\begin{equation}
\frac{|F_S(p_T^2, \mu_h) U_S(\mu_h, \mu)|}{|F_S^{\rm LO}(p_T^2, \mu)|} = 1.30\: (1 + 0.172 + 0.013)\,,
\end{equation}
where the three terms in the bracket correspond to LO, NLO and NNLO in RG-improved perturbation theory, which is equivalent to NLL, NNLL and N$^3$LL accuracy. However, both the corrections to $\widetilde{H}_{gg}$ and the (improved) scalar form factors $F_S$, as well as the ones to the jet and soft functions, happen to be positive. As a result, the expansion of the improved cross section is not much better behaved than the standard expansion. But given that all ingredients have well-behaved expansions and that the individual corrections may not necessarily add up constructively at higher orders, we are led to expect that the N$^3$LO corrections will be significantly smaller than the NNLO terms. For the form factors, the third-order corrections are known and indeed quite small \cite{Baikov:2009bg,Lee:2010cga,Gehrmann:2010tu}.

\begin{figure}[t!]
\centering
\includegraphics[width=0.5\textwidth]{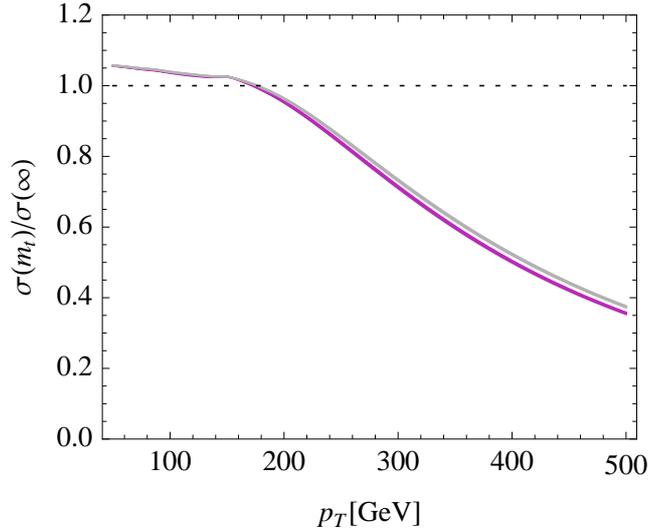} 
\caption{LO result at finite $m_t$ versus the result in the $m_t\to \infty$ limit. For the plot we have varied the scale in the range $p_T/2< \mu<2 p_T$ and have computed results for both $\sqrt{s}=8\,{\rm TeV}$ (purple) and $\sqrt{s}=13\,{\rm TeV}$ (gray). The resulting bands are very narrow and the ratio is also to very good accuracy independent of $\sqrt{s}$. 
\label{fig:mt}}
\end{figure}

Before proceeding to a detailed numerical analysis, we note that the hard function relevant for soft-gluon resummation of the total rate is given by the square of the scalar form factor $|F_S(\hat{s},\mu)|^2$. Our ansatz thus predicts that the rate for Higgs production with a jet suffers at large $p_T$ from the same corrections as the square root of the total rate. An alternative way to improve the predictions is thus to use the total cross section instead of the scalar form factor, when performing the improvement as in \eqref{eq:impr}. The RG-improved value of the total cross section can be obtained using the code {\sc RGhiggs} \cite{Ahrens:2010rs,rghiggs} and one needs to evaluate the cross section with $m_H$ set equal to $p_T$. Numerically, the results obtained in this way look quite similar to the improvement with the scalar form factor shown in Figure \ref{fig:scale}. 

In conclusion we find that the large perturbative corrections are associated with higher-order terms in the hard function. In principle, one can pursue a similar strategy as for the total cross section and use RG techniques to resum the corrections associated with the analytic continuation of the scalar form factor. We find, however, that even though the individual ingredients to the differential cross section have well-behaved perturbative expansions in such an approach, all terms happen to add up constructively and the NNLO correction to the cross section remains sizeable. Given the moderate improvement, we refrain from adopting this procedure when presenting numerical results for the spectrum in the next section.

\section{Numerical results\label{numer}}

 \begin{figure}[t!]
\centering
\begin{tabular}{ccc}
\includegraphics[width=0.30\textwidth]{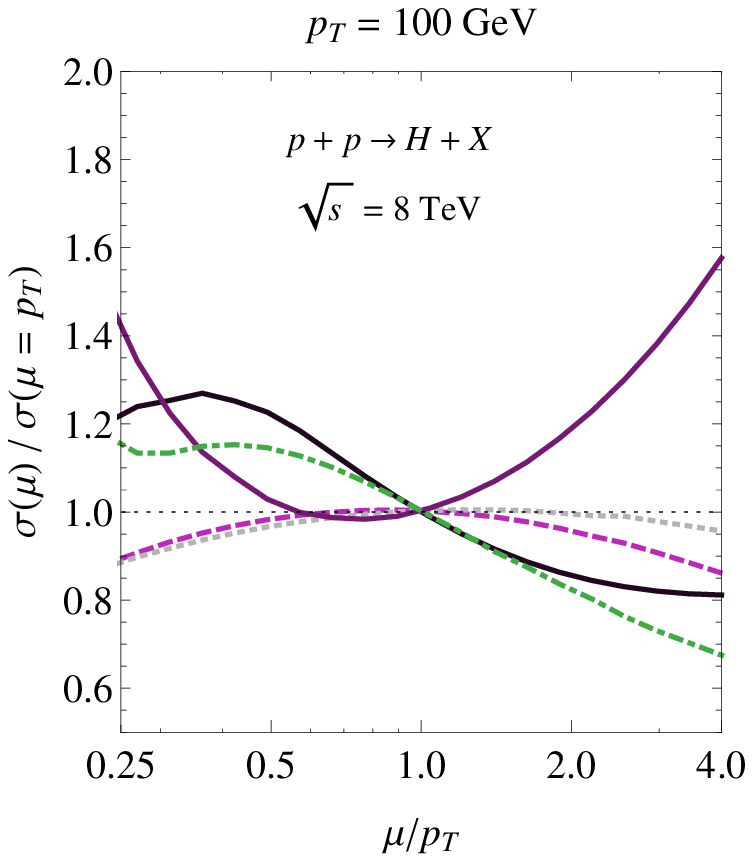} &
\includegraphics[width=0.30\textwidth]{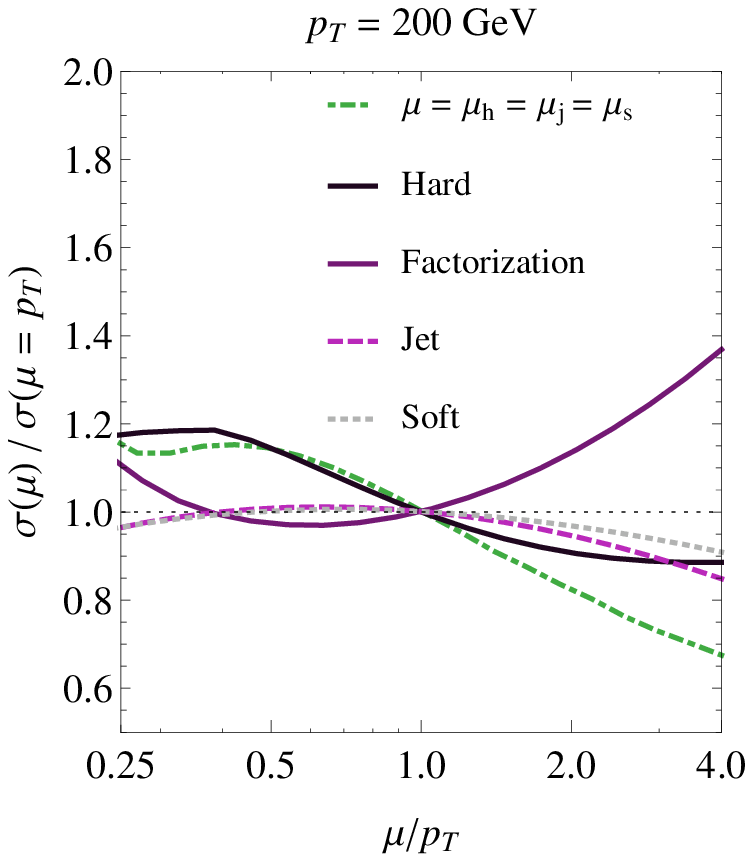} & 
\includegraphics[width=0.30\textwidth]{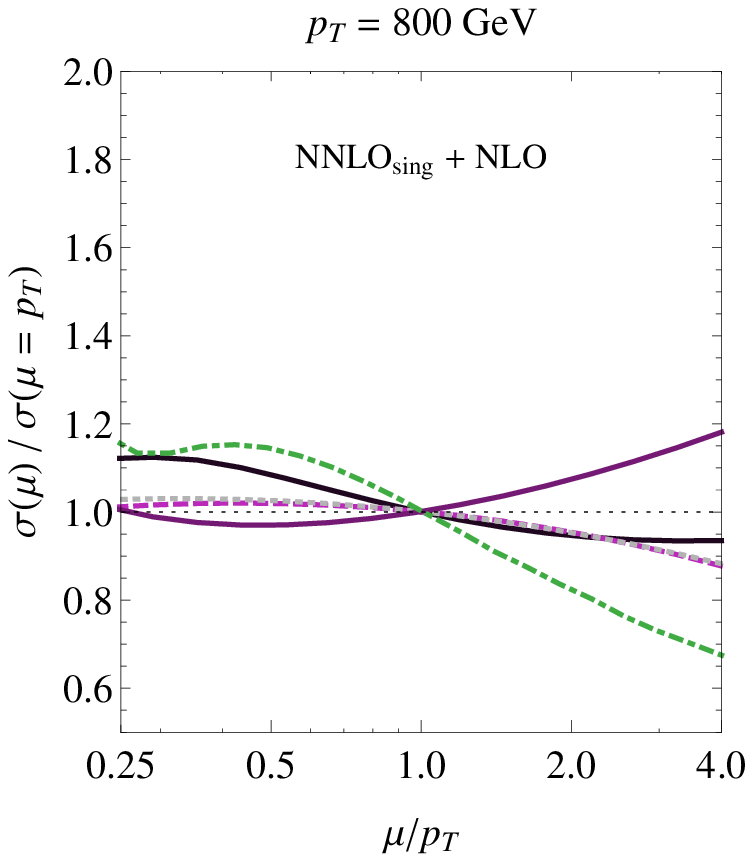} 
\end{tabular}
\caption{Individual scale variations of the cross section at different values of the transverse momentum.\label{fig:indiv}}
\end{figure}

Having discussed the size of the individual corrections, we now present numerical results for the transverse-momentum spectrum. For our predictions, we use MSTW2008NNLO PDFs \cite{Martin:2009iq} and their associated value for the strong coupling constant $\alpha_s(M_Z)=0.1171$. We further set $m_H=126\,{\rm  GeV}$ and $m_t = 173 \,{\rm GeV}$.

Before proceeding to the results, we need to discuss one important point. The factorization theorem \eqref{eq:factorization} holds both at finite $m_t$ and in the heavy top limit $m_t \to \infty$. However, the exact top-mass dependence has so far only been computed at leading order \cite{Ellis:1987xu,Baur:1989cm}. At NLO, one would need to compute two-loop four-point diagrams with massive top quarks, which is quite challenging. Our NNLO results for the hard function are therefore only valid in the heavy top limit, which is no longer adequate when the $p_T$ of the boson becomes of the order of the top quark mass. The exact leading order result has been implemented into the code {\sc HiggsPT} \cite{higgspt}. In Figure \ref{fig:mt} we show a comparison of the exact LO result with its $m_t \to \infty$ limit. The figure shows that for $p_T > 200\, {\rm GeV}$, the corrections to the heavy top limit become important. In the absence of the exact higher-order hard functions, the best way to take these effects into account is to multiply the higher-order results by the correction factor in Figure \ref{fig:mt}. We note that the factor is largely independent of the scale. The partonic cross section has identical scale dependence (given by the overall factor $\alpha_s(\mu)^3$ at LO), so that scale differences in the ratio only arise because the shape of the PDFs evolves when the scale is changed and they are integrated against a different weight in the numerator and denominator. The correction factor is also quite insensitive to the center-of-mass energy of the collider. In addition to the LO results,  the first order terms in an expansion in $1/m_t^2$ are known at NLO \cite{Harlander:2012hf}. This paper concluded that for $p_T < 200\, {\rm GeV}$ the NLO effects are not very large and that the bulk of the effects is captured by reweighting with the exact LO cross section, as discussed above. In addition to the finite quark mass effects, also electroweak corrections should be considered. Both types of corrections were analyzed in \cite{Keung:2009bs}, and it was found that also the electroweak effects are moderate below $p_T < 200\, {\rm GeV}$.

\begin{figure}[t!]
\centering
\begin{tabular}{cc} 
\includegraphics[width=0.48\textwidth]{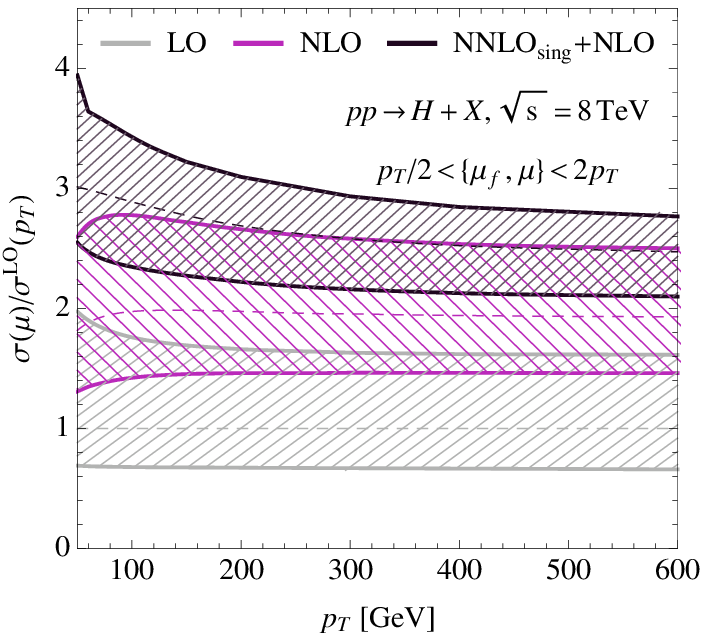} & \includegraphics[width=0.48\textwidth]{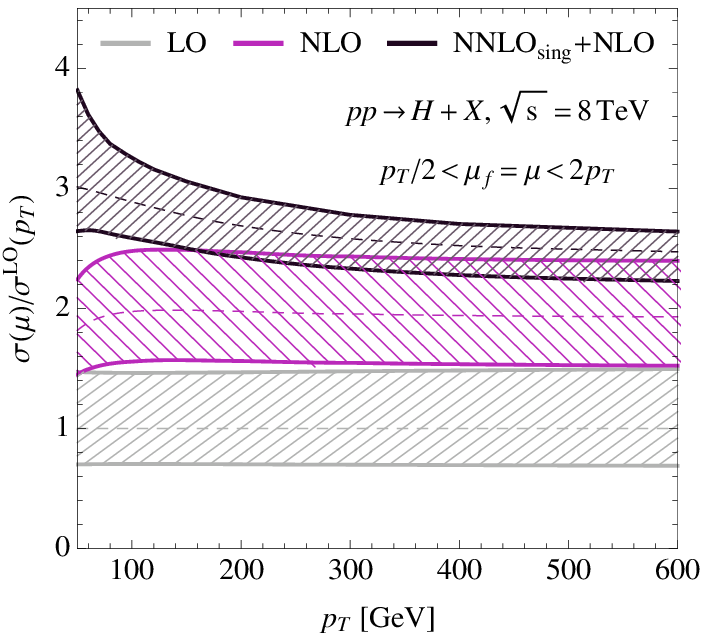}
\end{tabular}
\caption{Transverse-momentum spectrum at LO (gray), NLO (purple) and NNLO$_{\rm sing} +$NLO (black) at $\sqrt{s}=8\, {\rm TeV}$. Left: Independent variations of $\mu_f$ and $\mu=\mu_h=\mu_j=\mu_s$, see text. Right: Correlated scale variations $\mu_f = \mu$ by a factor of two.\label{fig:spectrum}}
\end{figure}

For our final results, we  use a conservative approach to estimate the size of missing higher-order corrections. We found in Section \ref{size} that there is no clear hierarchy between the jet, soft and hard scales, at least not at values of $p_T$ which are of phenomenological interest. We therefore do not perform any resummation, but simply set the different scales equal to a common scale $\mu$. However, in contrast to a standard fixed-order treatment, we can vary the scales separately in the different ingredients of our formula. The variation of the cross section from changing the hard, jet, soft and factorization scales individually is displayed in Figure \ref{fig:indiv}. The largest effects arise from the variation of the hard and factorization scales. For the hard scale, this is expected since the hard function receives the largest perturbative corrections. The factorization scale dependence provides an estimate of the missing non-threshold terms which would be needed to make the result independent of $\mu_f$ up to terms beyond NNLO. The large $\mu_f$ dependence at smaller $p_T$ indicates that non-threshold corrections could play an important role in this region. From Figure \ref{fig:indiv}, we observe that the variations of $\mu_h$ and $\mu_f$ tend to go in opposite directions. Varying the two scales together might therefore not provide a reliable uncertainty estimate, and we vary the scales both individually and in a correlated way. Specifically, we set $\mu=\mu_h=\mu_j=\mu_s$ and vary $\mu$ and $\mu_f$ separately up and down by factors of two around the default value $\mu=\mu_f =p_T$, while constraining $1/2 \leq \mu_f/\mu < 2$. This yields seven values for the cross section at a given value of $p_T$ and we define the scale uncertainty band by the maximum and minimum values. From Figure \ref{fig:indiv}, we  observe that  the $\mu_f$ variations decrease at higher $p_T$ values, which arises because the threshold contributions become more dominant.

\begin{figure}[t!]
\centering
\begin{tabular}{cc} 
\includegraphics[width=0.495\textwidth]{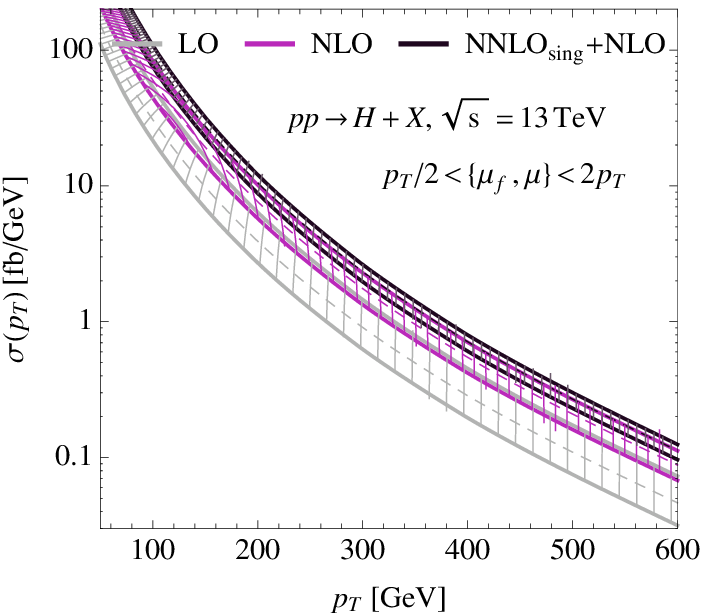} & 
\includegraphics[width=0.48\textwidth]{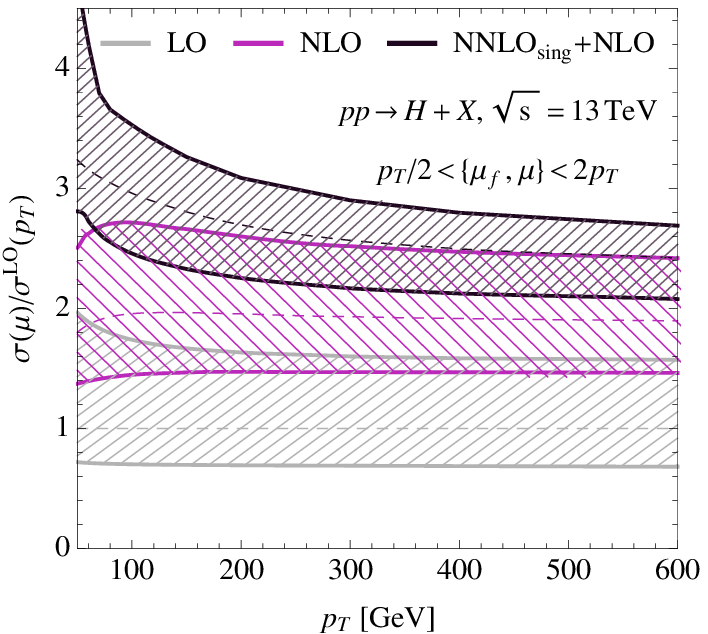}
\end{tabular}\caption{Transverse-momentum spectrum at LO (gray), NLO (purple) and NNLO$_{\rm sing} +$NLO (black) at $\sqrt{s}=13\, {\rm TeV}$.\label{thirteen}}
\end{figure}

The result of the uncorrelated scale variation is shown in the left panel of Figure \ref{fig:spectrum}. The largest variations are due to the individual $\mu$ or $\mu_f$ variations, which are explicitly given in Table \ref{tab:values}. The lower boundary of the scale bands always arises from varying $\mu$ upward, while the upper boundary is set by different variations, depending on the value of $p_T$. The kink in the upper edge of the NNLO band near $p_T=60\,{\rm GeV}$, for example, arises because the maximum switches to a different variation at that point. For comparison we show in the right panel the scale bands obtained from a correlated variation of $\mu =\mu_f$ by a factor of two. In contrast to the more conservative approach we use here, the bands do not fully overlap with this prescription. We adopt the more conservative prescription to present the results for $\sqrt{s}=13\,{\rm TeV}$ in Figure \ref{thirteen}. In Table \ref{tab:values}, we present values for the cross section and  the separate $\mu$ and $\mu_f$ variations. When computing individual scale variations, a small amount of resummation is being performed because the RG-evolution factors in our resummed result become nontrivial (their explicit form was given in \cite{Becher:2013vva,Becher:2012xr}). For this reason, we need to distinguish NLO$_{\rm sing}$+NLO from standard NLO. In the NLO$_{\rm sing}$+NLO result, the perturbative corrections to the hard, jet and soft functions are evaluated at the scale $\mu$ and the result is RG evolved to the scale $\mu_f$, where the matching corrections are added and the convolution with the PDFs is performed. In contrast, in fixed-order computations the dependence on the renormalization scale $\mu$ is obtained by starting with the perturbative result computed with a single scale $\mu_f$ and then reexpanding in terms of a coupling constant at a different scale $\mu$. As the entries in Table \ref{tab:values} show,
our prescription leads to a more conservative error estimate. In the table, we also give PDF and $\alpha_s$ uncertainties. To obtain those, we have used the MSTW2008NNLO 90\% confidence level error PDFs and the associated $\alpha_s(M_Z)=0.1171\pm 0.0034$. The uncertainties are given for our highest-order result, but the relative uncertainties are largely independent of the order if the same PDF set is used. The uncertainty on $\alpha_s$ is larger than the PDF uncertainty and to a good approximation simply a result of the overall $\alpha_s^3$ prefactor.

We finally briefly compare our numbers to the NNLL results of \cite{Huang:2014mca}. This paper found that higher-order corrections lower the cross section, while we find a large increase at two-loop order. The reason for this difference is that the dominant corrections come from the two-loop hard function, which is not included in the result of \cite{Huang:2014mca}. We have tried to numerically compare results at NNLL accuracy, but the fact that the authors only show plots and do not fully specify how the uncertainty bands are generated makes a detailed comparison difficult. Adopting the same default scale choices as \cite{Huang:2014mca}, we find results which appear to be consistent with the plots in this paper. We note that \cite{Huang:2014mca} uses fixed values for the jet and soft scales, while the plots in Figure \ref{fig:hjsH} seem to indicate that they scale with the transverse momentum. Also, the value of the hard scale $\mu_h = 2.5 \,\sqrt{p_T^2+m_H^2}$ adopted in \cite{Huang:2014mca} is quite high.

\begin{table}[t!]
\centering
\begin{tabular}{lllll}\toprule
\multirow{2}{*}{$\frac{d\sigma}{dp_T}\,[{\rm fb}/{\rm GeV}]$} &  \multicolumn{2}{c}{LHC at $8\,$TeV}
& \multicolumn{2}{c}{LHC at $13\,$TeV}
\\ 
& $100\,{\rm GeV}$ &  $200\,{\rm GeV}$ &  $100\,{\rm GeV}$ &  $200\,{\rm GeV}$
\\ \midrule
LO$_{\rm sing}$ &
${11.0}_{-3.5-0.4}^{+8.4+1.8}$ & ${1.17}_{-0.38-0.05}^{+0.77+0.13}$ & 
${\phantom{1}30.6}_{\phantom{1}-9.1-1.2}^{+22.6+5.9}$ & ${\phantom{1}3.90}_{-1.19-0.16}^{+2.47+0.50}$ 
\\ 
NLO$_{\rm sing}$ 
& 
${25.1}_ {-6.1-0.3}^{+8.8+3.8}$ &    ${2.48}_ {-0.6-0.07}^{+0.8+0.28}$ &  
${\phantom{1}71.2}_ {-15.6-0.6}^{+23.2+11.5}$ &     ${\phantom{1}8.34}_ {-1.89-0.19}^{+2.52+1.02}$
\\ 
NNLO$_{\rm sing}$ &
${35.2}_ {-6.0-0.3}^{+5.3+5.3}$ &     ${3.31}_ {-0.55-0.11}^{+0.48+0.41}$ &    
${101.8}_ {-15.3-0.1}^{+13.7+15.3}$ &    ${11.23}_ {-1.73-0.23}^{+1.52+1.47}$ 
\\ \midrule
NLO & ${21.7}_{-4.1-0.6}^{+5.0+0.6}$ & ${2.31}_{-0.39-0.09}^{+0.48+0.11}$ & 
${\phantom{1}59.8}_{-11.1-0.6}^{+13.5+0.6}$ & ${\phantom{1}7.63}_{-1.29-0.19}^{+1.55+0.23}$
\\
NLO$_{\rm sing}$+NLO &
${21.7}_ {-6.1-0.6}^{+8.8+4.4}$ &    ${2.31}_ {-0.60-0.06}^{+0.80+0.29}$ &      
${\phantom{1}59.8}_ {-15.6-1.4}^{+23.2+13.6}$ &   ${\phantom{1}7.63}_ {-1.89-0.17}^{+2.52+1.07}$ 
\\
NNLO$_{\rm sing}$+NLO & 
${31.8}_ {-6.0-0.5}^{+5.3+5.9}$ &     ${3.14}_ {-0.55-0.09}^{+0.48+0.42}$ &    
${\phantom{1}90.5}_ {-15.3-0.5}^{+13.7+17.4}$ &   ${10.52}_ {-1.73-0.21}^{+1.52+1.53}$ 
\\ \midrule
PDF uncertainty & $\phantom{31.5}^{+3.2\%}_{-3.4\%}$ & $\phantom{31.5}^{+4.3\%}_{-4.4\%}$ & 
$\phantom{31.5}^{+2.4\%}_{-2.7\%}$ & $\phantom{31.5}^{+3.1\%}_{-3.3\%}$
\\[0.1em]
$\alpha_s$ uncertainty & $\phantom{31.5}^{+11.9\%}_{-10.9\%}$ & $\phantom{31.5}^{+10.4\%}_{-9.7\%}$ & 
$\phantom{31.5}^{+11.9\%}_{-10.9\%}$ & $\phantom{31.5}^{+10.4\%}_{-9.7\%}$ \\ \midrule
$d\sigma^{\rm LO}(m_t)/d\sigma^{\rm LO}(\infty)$ & 1.036 & 0.954 & 1.039 & 0.964
\\
\bottomrule
\end{tabular}
\vspace{0.3cm}
\caption{Results for the cross section and its scale uncertainty using different approximations, see text. The scale uncertainties are obtained by  varying the scales $\mu=\mu_h=\mu_j=\mu_s$ and $\mu_f$ by a factor of two around the default value $\mu=\mu_f=p_T$. The first uncertainty is the variation of $\mu$, the second one $\mu_f$. \label{tab:values}}
\end{table}

\section{Conclusion}

We computed the NNLO corrections to the Higgs transverse-momentum spectrum in the threshold limit. The threshold corrections as well as the N$^{3}$LL resummed results are implemented in the public code {\sc PeTeR} \cite{peter}. The NNLO corrections turn out to be sizeable, and we gave a detailed discussion about the origin of these corrections. Similar to the inclusive Higgs production cross section, they are associated with higher-order terms in the hard function and can be resummed using RG techniques. The RG improvement turns out, however, to be not very efficient for the transverse-momentum spectrum.

Our analysis revealed that there is no pronounced hierarchy between the hard, jet and soft scales, and we thus refrained from resumming the threshold terms to all orders. However, we used the  scale separation to obtain a more conservative uncertainty estimate than in fixed-order calculations which seems appropriate in view of the large corrections. The dominance of the virtual corrections further implies that the threshold expansion should provide a good approximation of the full NNLO result even at moderate values of $p_T$. 

Our result will serve as a check of the full NNLO Higgs plus one jet calculation once it becomes available. It includes all partonic channels, and it turns out that the $q\bar{q}$ channel is negligible but the $qg$ contribution is numerically significant. Our calculation also provides an estimate of beyond NNLO corrections. The dominant N$^3$LO terms will likely arise in the hard function, and they can be estimated using the improvement scheme introduced in Section \ref{size}.

Interestingly, we find that the NNLO terms lead to changes in the shape of the $p_T$ distribution which are comparable in size to finite-$m_t$ effects. This could be relevant, for example, in the context of new physics searches at large transverse momentum using methods such as the ones advocated in \cite{Banfi:2013yoa,Azatov:2013xha,Grojean:2013nya,Schlaffer:2014osa}. Preliminary experimental results for the Higgs transverse-momentum spectrum are already available \cite{TheATLAScollaboration:2013eia}. These measurements are based on the decay $H\to\gamma\gamma$, and they reach up to transverse momenta of about $200\,{\rm GeV}$. The higher energy and luminosity of Run II will allow to extend the measurements to higher $p_T$ values, in particular if also larger decay channels such as $H\to\tau\tau$ are taken into account. We look forward to comparing our results to these measurements.

\vspace{0.6cm}
{\em Acknowledgments:\/}
We thank Massimiliano Procura for useful discussions on the structure of the corrections to the hard function. 
We thank Fabrizio Caola, Frank Petriello and Markus Schulze for pointing out a numerical problem in the quark-gluon channel in the preprint version of this paper and Thomas Gehrmann and Matthieu Jacquier for providing a corrected helicity amplitude for this channel. The work of T.B.\ is supported by the Swiss National Science Foundation (SNF) under grant 200020-140978. G.B.\ gratefully acknowledges the support of a University Research Fellowship by the Royal Society.


\end{document}